# Enhanced Acoustic Beamforming with Sub-Aperture Angular Multiply and Sum - in vivo and in Human Demonstration

Matthieu Toulemonde†, Cameron A. B. Smith†, Kai Riemer, Priya Palanisamy, Jaideep Singh Rait, Laura Taylor, Peter D. Weinberg, Karina Cox, and Meng-Xing Tang

*Abstract*— Power Doppler ultrasound is in widespread clinical use for non-invasive vascular imaging but the most common current method - Delay and Sum (DAS) beamforming - suffers from limited resolution and high side-lobes. Here we propose the Sub-Aperture Angular Multiply and Sum (SAMAS) algorithm; it combines the advantages of two recent non-linear beamformers, Frame Multiply and Sum (FMAS) which uses signal temporal coherence and the acoustic sub-aperture (ASAP) algorithm, which uses signal spatial coherence, respectively. Following *in vitro* experiments to optimise the algorithm, particularly the use of phase information and sub-aperture pairing, it was evaluated *in vivo*, first in a rabbit kidney and then in human lymph node, using ultrafast ultrasound images obtained with intravenous contrast agents. The SAMAS algorithm improved the CNR and SNR across all tests, on average raising the CNR by 11 dB and the SNR by 18 dB over DAS in vivo. This work demonstrates a promising vascular imaging method that could have widespread clinical utility.

## I. Introduction

POWER Doppler (PD) ultrasound is a non-invasive method for imaging vascular perfusion, which it achieves by detecting the movement of red blood cells from Doppler shifts [1]. PD's ability to measure vascularity is frequently used as a clinical tool for identifying presence of flow, detecting decreased flow that is characteristic of areas of ischemia [2], and demonstrating inflammatory hyperaemia or irregular blood flow in tumours.

Ultrafast PD utilises high frame rate ultrasound images to generate the PD signal over a large field of view [3], [4]. In order to compute the ultrafast PD images, tissue has to be filtered from the beamformed data in order to isolate the moving blood signals. Besides frequency domain filtering, singular value decomposition (SVD) is increasingly used for this purpose [5], [6].

Beamforming is a key step in generating these images. The most commonly used method is the delay and sum (DAS) algorithm with Coherent Compounding (CC) [7], [8], [9]. However, it suffers from limited resolution and high side-lobes, limiting its effectiveness [10].

Several methods have recently been developed to enhance ultrafast doppler imaging. They include Short-Lag Spatial Coherence [11], Cross-Angular Delay Multiply and Sum [12], minimum variance beamforming [13], [14], and temporal coherence based estimators [15]. Of particular relevance to this study is the development of the frame multiply and sum (FMAS) [16] and the acoustic sub-aperture (ASAP) [17] methods. The FMAS method capitalizes on the temporal coherence of signals across different transmitted plane wave angles: beamformed radio frequency (RF) signals from each of the transmitted angles are cross-multiplied together before being averaged, thereby taking advantage of the correlation between each to improve the contrast and resolution over traditional DAS.

In the ASAP method, contrast is improved by splitting the receive aperture in half, generating two images each comprising information from half of the available elements; they therefore contain common back scattered signals but different electronic noise. The two images are then multiplied and averaged over time to generate an image based on the correlation between them enhancing vascular flow signals and reduce noise. Potential grating lobe artefacts due to increased distance between the elements during beamforming, which can lead to grating-lobe artefacts. These grating lobes can be suppressed by suppressing pixels that contain negatively correlated phases [17].

Ultrasound contrast agents consist of gas filled particles which, when present in blood, provide high contrast compared to surrounding tissue due to their sharp acoustic mismatch. The most common type of ultrasound contrast agents are microbubbles, which are clinically approved and have therefore seen widespread use [18]. Contrast enhanced ultrasound provides a highly sensitive approach for imaging the vasculature through which the microbubbles flow, and has proven to be valuable in imaging a wide range of organs including the liver, heart, kidney, carotid artery, thyroid nodules, pancreas, and breast, characterising

†These authors contributed equally to this work. The corresponding author is Meng-Xing Tang. The authors gratefully acknowledge funding from and the National Institute for Health Research i4i (grant number NIHR200972), the International Human Frontier Science Program Organization (grant LT0036/2022-L to C.A.B.S.) and Engineering and Physical Sciences Research Council (EPSRC) under Grant EP/T008970/1. .M. Toulemonde, C.A.B. Smith, L. Taylor, P. D. Weinberg, and M.X. Tang, are with the Department of Bioengineering, Imperial College London,SW7 2AZ London, U.K. (e-mail: m.toulemonde@imperial.ac.uk; cameron.smith13@imperial.ac.uk; laura.taylor19@imperial.ac.uk;p.weinberg@imperial.ac.uk; mengxing.tang@imperial.ac.uk).K. Riemer is with the Institute of Mountain Emergency Medicine, 39100 Bolzano, Italy (email: kai.riemer@eurac.edu).P. Palanisamy, Jaideep Singh Rait, and K. Cox, are with the Maidstone and Tunbridge Wells NHS Trust, ME16 9QQ, Maidstone, U.K. (email:p.palanisamy@nhs.net; j.rait@nhs.net; karina.cox@nhs.net)

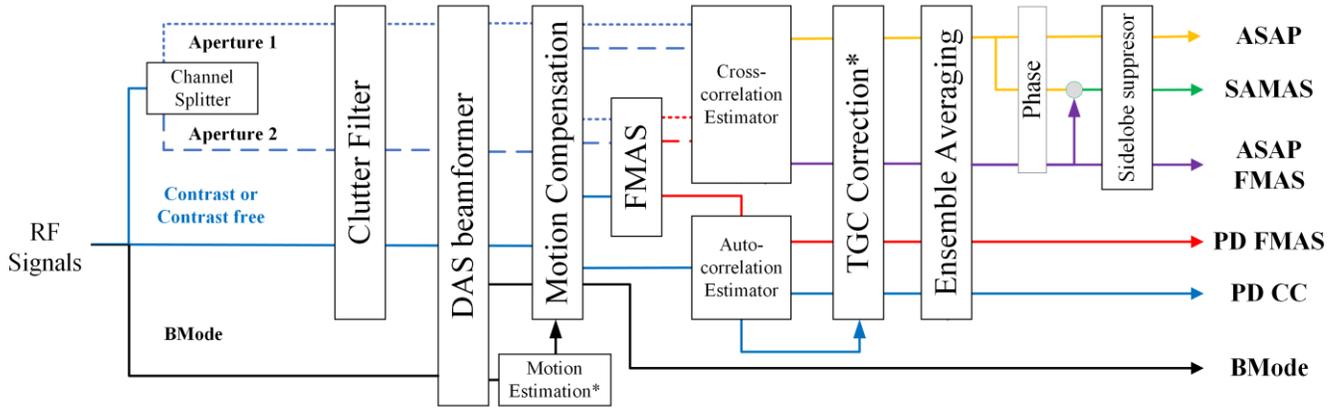

Fig. 1. Alternative signal processing paths of Power Doppler (PD) Delay-And-Sum (DAS), PD Frame-Multiply-and-Sum (FMAS), Acoustic Sub-APerture (ASAP), ASAP FMAS and Sub-aperture Angular Multiply and Sum (SAMAS) imaging for contrast and contrast free acquisition. * motion estimation and Time Gain Compensation (TGC) are only applied on *in-vivo* acquisition.

tumours [19], [20] and evaluating perfusion [21], [22].

In this work we describe a comparison of FMAS and ASAP beamforming techniques in the context of ultrafast contrast enhanced ultrasound and propose an enhanced method that combines them, which we term Sub-aperture Angular Multiply and Sum (SAMAS), to gain greater image enhancement.

## II. MATERIALS AND METHODS

Five techniques are presented in this work, PD CC, PD FMAS, ASAP, combined ASAP-FMAS, and the Sub-aperture Angular Multiply and Sum (SAMAS). First, an in-vitro dataset is gathered to allow for the generation of optimised imaging parameters. Following this optimisation two in vivo datasets are explored in order to evaluate these algorithms, one of a rabbit kidney, and the other of human lymph nodes in a clinical setting.

*A. Techniques*

In this study, five techniques are evaluated: PD CC, PD FMAS, ASAP, ASAP FMAS, and SAMAS. A diagram demonstrating each technique is shown in Figure 1.

One in vitro and two in vivo datasets are generated as detailed in the following sections. In the case of our *in vivo* datasets, after the acquisition of the raw datasets, received RF signals for all of the elements are processed in two different ways. In the first case we beamform the data as a traditional CC B-Mode image, which is then fed into the 'imregdemons' function of Matlab (MathWorks, Natick, MA, USA) in order to produce an estimate of the motion fields in the data. Then in the main processing pipeline we apply a clutter filter to the RF signals, either SVD (for the kidney images) or temporal filter (for the others), before beamforming occurs. We then conduct motion correction utilising the motion field derived previously. The motion corrected CC images are then placed into an auto-correlation estimator in order to produce parameters to correct the time gain compensation (TGC). These TGC parameters and motion fields are applied to all the other beamforming methods to ensure that that any enhancements seen in the final images are not due to changes in motion correction or TGC correction, thereby ensuring a fair comparison in the final images.

*1) Power Doppler (PD) Coherent Compounding (CC)*

PD CC imaging is most commonly performed using the DAS beamforming method [9]. The PD CC images are then generated from the autocorrelation of these images after clutter filtering.

*2) PD Frame Multiply And Sum (FMAS)*

In FMAS [16] images produced from each angular transmission are cross multiplied before summation. This is given by

$$y_{FMAS}(x) = \sum_{a=1}^{A-1} \sum_{b=a+1}^{A} sign(y_{DAS,a} y_{DAS,b}) |y_{DAS,a} y_{DAS,b}| \quad (1)$$

This technique has shown to enhance image resolution and contrast, and to reduce sidelobe artefacts.

*3) Acoustic Sub-APerture (ASAP)*

The ASAP algorithm [17] enhances image contrast by splitting the receive aperture in two halves, generating two images, each using half of the receiving elements. These two images after clutter filtering are then multiplied and summed over multiple frames, generating an image based on the correlation between them. This aperture splitting can be done in a variety of ways, this can include taking every other element [1 0 1 0 1 0…] denoted [1 0] for short; and taking pairs of elements [1 1 0 0 1 1 0 0…] denoted [1 1 0 0] and so on.

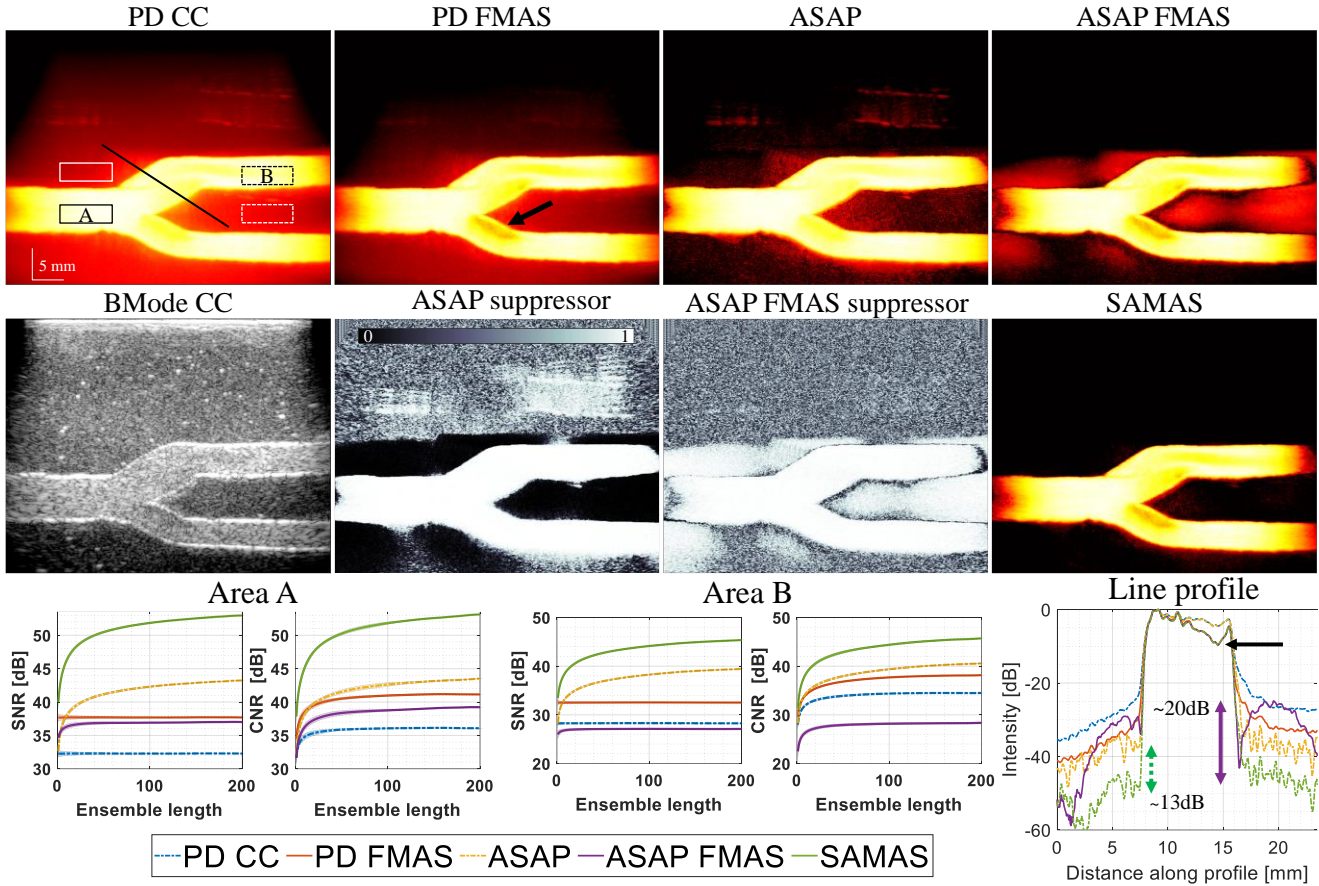

Fig. 2. BMode, Power doppler CC, FMAS, ASAP, ASAP FMAS and SAMAS images of a carotid bifurcation phantom are displayed with a 50db dynamic range. Sidelobe suppressor coefficients for ASAP and SAMAS and ASAP FMAS are displayed with 0 and 1. The SNR and CNR over varying ensemble lengths for areas A and B and a cut-through line profile comparing each of the beamforming techniques can also be seen.

TABLE I
SIGNAL-TO-NOISE RATIO (SNR), AND CONTRAST-TO-NOISE RATIO (CNR) FOR TWO ROIS AS INDICATED IN FIG 2 FOR AN ENSEMBLE LENGTH OF 200 SAMPLES. THE BOLD TYPE PROVIDES THE dB DIFFERENCE BETWEEN SAMAS AND OTHER METHODS (POSITIVE VALUES MEANS IMPROVEMENT).

|  | ROI | PD CC | PD FMAS | ASAP | ASAP FMAS | SAMAS |
|---|---|---|---|---|---|---|
| SNR (dB) | A | 32.3 (20.7) | 37.7 (15.3) | 43.2 (9.8) | 37.0 (16.0) | **53.0** |
| SNR (dB) | B | 28.3 (17.1) | 32.5 (12.9) | 39.4 (5.9) | 27.1 (18.3) | **45.4** |
| CNR (dB) | A | 36.1 (17.1) | 41.1 (12.0) | 43.5 (9.6) | 39.2 (13.9) | **53.1** |
| CNR (dB) | B | 34.5 (11.2) | 38.1 (7.6) | 40.6 (5.1) | 28.3 (17.4) | **45.7** |

One downside of splitting the aperture in this way is the introduction of grating lobes artefacts. This is caused by larger transducer pitch size and results in negatively correlated interference signals. These artifacts can then be reduced by applying a suppressor, a weight vector determined by the phase of the correlated signal, in which negatively correlating pixels are suppressed.

*4) Sub-aperture Angular Multiply and Sum (SAMAS)*

In the proposed method we utilise the advantages of both ASAP and FMAS. However, if we directly perform ASAP and FMAS together the cross-multiplication makes the negative phase correlations which are utilised as part of the side-lobe suppression into positive correlations, making them impossible to discriminate from the main lobe, introducing strong artefacts into the image. However, if instead we use the phase information from the standard ASAP and use negative phase correlations from that to determine the side-lobe suppressor, we can then apply these weight values to the ASAP FMAS algorithm to produce our improved SAMAS images. The additional computational cost involved in this step is negligible as ASAP images can be generated from the same beamformed dataset, so no additional beamforming is required.

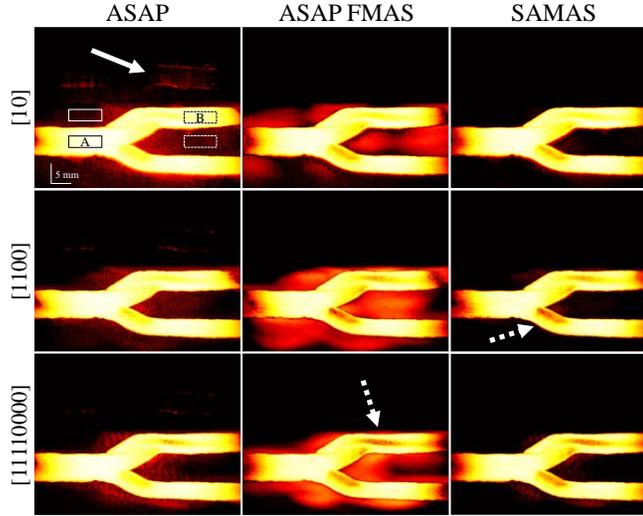

Fig. 3. Power doppler ASAP images. 50db Dynamic range. Exploring the optimal pairings of elements as part of the ASAP method, pairing adjacent elements [1 0] to groups of four elements [1111 0000].

TABLE II
SIGNAL-TO-NOISE RATIO (SNR) AND CONTRAST-TO-NOISE RATIO (CNR) FOR TWO ROIS AS INDICATED IN FIG 4 FOR AN ENSEMBLE LENGTH OF 200 SAMPLES FOR DIFFERENT ASAP APERTURE SHAPE.

|  |  | [10] |  |  | [1100] |  |  | [11110000] |  |  |
|---|---|---|---|---|---|---|---|---|---|---|
|  | ROI | ASAP | ASAP FMAS | SAMAS | ASAP | ASAP FMAS | SAMAS | ASAP | ASAP FMAS | SAMAS |
| SNR (dB) | A | 43.2 | 37.0 | **53.0** | 38.8 | 24.3 | 44.5 | 42.4 | 35.6 | 50.6 |
|  | B | 39.4 | 27.1 | **45.4** | 35.7 | 16.5 | 38.6 | 39.6 | 28.9 | 44.6 |
| CNR (dB) | A | 43.5 | 39.2 | **53.1** | 38.3 | 23.8 | 42.8 | 41.2 | 33.5 | 48.8 |
|  | B | 40.6 | 28.3 | **45.7** | 35.8 | 17.9 | 38.6 | 40.4 | 30.1 | 44.6 |

### B. Imaging platform

For the creation of all three of our datasets, a Vantage-256 Verasonics platform (Verasonics Inc. Redmond, WA) with a mounted a GE L3-12-D transducer (GE L3-12-D, GE HealthCare, Chicago, IL, USA) having 256 multiplexed elements with a pitch of 0.200 μm and a central frequency of 6.5 MHz was used. As all elements cannot be accessed at the same time, a one-aperture mode, using only the 128 central elements, or two-aperture mode, using two apertures of 128 elements with an overlap of 16 elements, have been developed for *in vitro* and *in vivo* acquisitions.

Non contrast B-Mode and contrast enhanced three pulse Amplitude Modulation (AM) with a 5 MHz half cycle (electrically, which produce a 3-cycle pulse as measured by hydrophone) transmission were used for the experiments. AM was implemented by emitting three coded pulses sequentially, two of which were transmitted with half the amplitude of the third (Full amplitude) pulse, Half-Full-Half, for each angle of each active aperture. As the Verasonics power system is not linear, half amplitude is generated by using half of the elements of the activate aperture for the first half. The second half is obtained using the previously unused elements. All the elements of the same aperture are used in reception.

The performance of the proposed algorithm was investigated though a series of *in vitro* and *in vivo* experiments in an animal model, before exploring the translatability of this method to the clinic by exploring an *in vivo* human application, detailed below.

### C. In Vitro Experimental Setup

The first aim was to test the feasibility of SAMAS and optimise the aperture splitting. An anatomically realistic polyvinyl alcohol (PVA) carotid bifurcation phantom as described in Lai *et al.* [23] was used to evaluate the proposed beamforming methods and explore potential optimisations *in vitro*. A gravity flow setup was employed to create laminar flow with a flow rate of 200 cc/min using water and in-house microbubbles with a concentration of 0.10 ml/L. These in-house manufactured microbubbles are lipid-shelled decafluorobutane bubbles, the manufacturing of which has been described previously [27]. For the ultrasound acquisition, a two-aperture AM acquisition was used with 10 angles over a 20-degree range with a pulse repetition frequency of 10 kHz with a Tukey 0.25 window used in transmission generating an MI of 0.10. 200 frames were acquired at 150 frames/s.

### D. In Vivo Rabbit Experimental Setup

The experiment complied with the Animals (Scientific Procedures) Act 1986 and received approval from the Animal Welfare and Ethical Review Body of Imperial College London under Project License No. P15180DF2. A pathogen-free male New Zealand White rabbit (HSDIF strain, Envigo, Huntingdon, UK) aged 20 weeks and weighing 3.14 kg was maintained on a 12:12 h light:dark cycle at 18°C and fed a standard laboratory diet. It was pre-medicated with acepromazine (0.5 mg/kg, i.m., CVet, London, UK), anaesthetised using a combination of medetomidine (Domitor, 0.25 mL/kg, i.m.) and ketamine (Narketan, 0.15 mL/kg, i.m.) and maintained with a third of the initial dose of anaesthetics every 30 to 45 minutes for up to 4 hours. The rabbit underwent tracheotomy and was ventilated at 40 breaths/minute. It was positioned supine on a heated mat with, continuous monitoring of heart rate and blood oxygen saturation, and fur was removed from the abdominal region. Euthanasia employed pentobarbital (0.8 mL/kg).

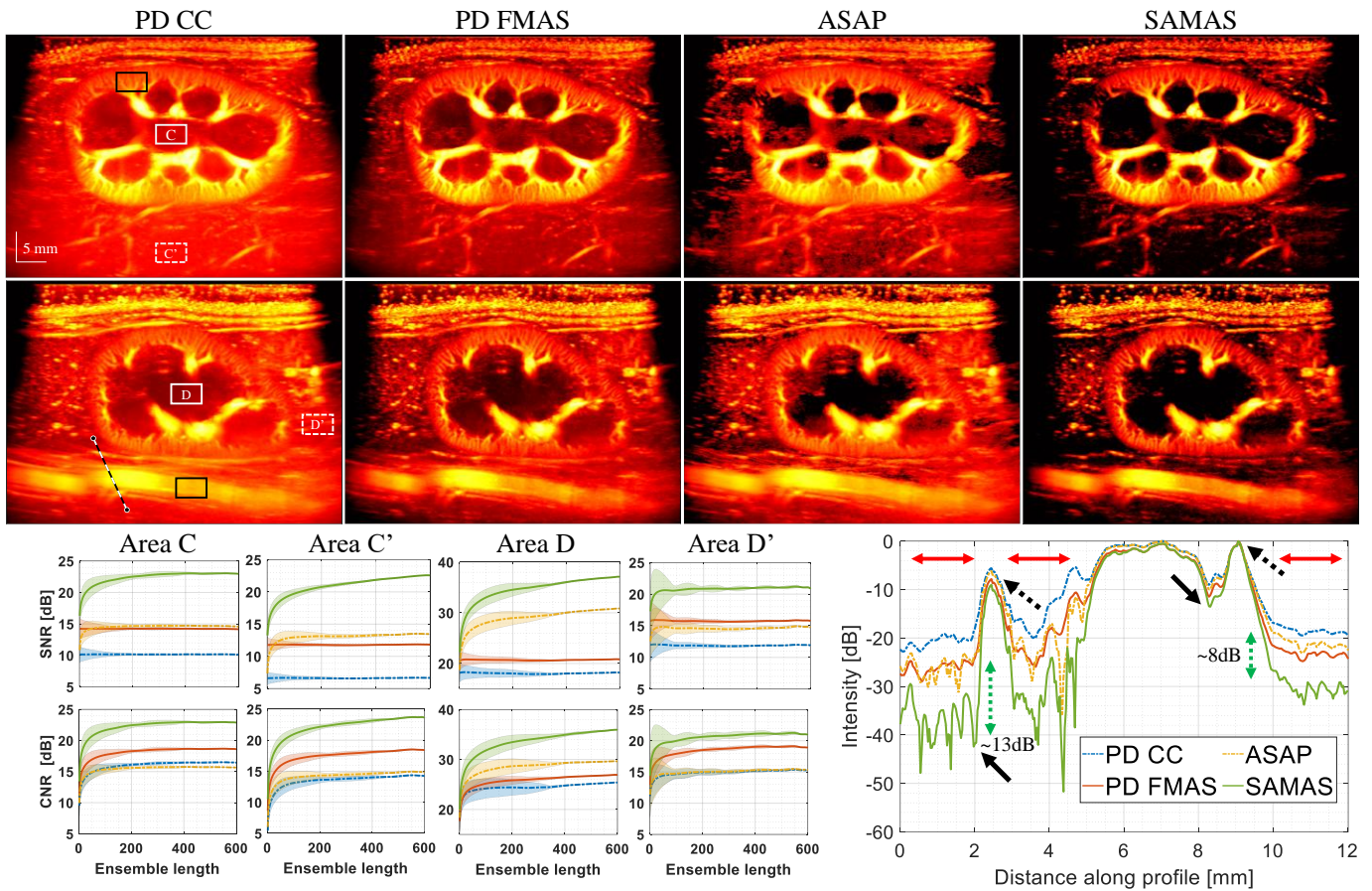

Fig. 4. Power doppler CC, FMAS, ASAP, and SAMAS images of blood signal from a white rabbit kidney displayed with a 45dB Dynamic range. The SNR and CNR over varying ensemble lengths for areas D and D'. A cut-through line profile comparing each of the beamforming techniques can also be seen.

We imaged the kidney transcutaneous using a two-aperture BMode acquisition, 0.66 MI, 10 angles over a 10-degree range and 300 frames per second. Two different views were acquired with the first view focused on the kidney and its cortex part while the second view is focused on the kidney and the aorta below.

### E. In Vivo Lymph Acquisition

The human acquisition is part of the LiSENUS -Lymphatic imaging Sentinel Node Super-Res CEUS project. Approval of the human ethical and experimental procedures and protocols was granted by the Research Ethics Committee (REC) of the North East - Tyne & Wear South Research Ethics Committee under the REC reference 21/NE/0150.

For the *in vivo* human lymph acquisition a bolus injection of 1.25 mL SonoVue (Bracco, Milan, Italy) was used, acquisitions were acquired after waiting 10s. For the ultrasound acquisition, one aperture mode transmission was used with 10 angles over a 20-degree range with a pulse repetition frequency of 7 kHz with a Tukey 0.25 window used in transmission generating an MI of 0.05. 1000 frames were acquired at 100 frames/s.

### F. Image processing and Analysis

To extract or improve flow information, two clutter filters have been used: SVD and rolling subtraction. SVD processing was applied for blood flow imaging and was done in Matlab using the option *'econ'* during the decomposition. Thresholds were selected using the L-curve described by Kang *et al.* [16]. The rolling subtraction was applied to the contrast RF data, and it acts as a high pass filter and removes residual static tissue. Both clutter filters are used before aperture splitting and beamforming of each individual angles to provide the same flow dataset to each beamforming. Based on the frequency response of the moving average filter, temporal high pass filtering, the half-power (-3 dB) cut-off frequency is 5.0 Hz for the lymph acquisition. Using the Doppler frequency formula, given a transmission frequency of 5 MHz and an acquisition frame rate of 100 fps, the axial cut-off velocity for this window is 0.8 mm/s. *In-vivo* dataset may suffer to a non-optimal TGC setting during the acquisition so a TGC compensation is conducted in a similar fashion to Song *et al.* [24].

For image analysis both CNR and SNR were calculated. The equations for which can be found below:

$$SNR = \frac{\mu_A}{\mu_B} \quad (2)$$

Where

$$CNR = \frac{\mu_A - \mu_B}{\sigma_B} \quad (3)$$

## III. RESULTS

### A. In Vitro Experimental Results

To check the effectiveness of the SAMAS algorithm we initially used an *in vitro* bifurcation phantom, where Contrast, B-mode as well as phase information for two techniques are shown in Figure 2. A line cut through of the intensities across the phantom and the mean and standard deviation of the CNR and SNR as a function of ensemble length at two locations are also provided. Qualitatively it can be seen that the SAMAS algorithm has improved contrast and artefact suppression when compared to PD CC, PD FMAS, ASAP, and ASAP FMAS. The line cut and the graphs also demonstrate the improved CNR and SNR of SAMAS but also demonstrates how that this improvement continues to improve at increasing ensemble lengths, gaining further SNR and CNR in the same way that conventional ASAP imaging does. SNR and CNR values at an ensemble length of 200 can be seen in Table I.

It can be seen that artefacts due to the multiple reflections are reduced due to the FMAS aspects of the algorithm, however the reduction in sidelobes is due to the SAMAS combination.

We then explored optimising the SAMAS algorithm, looking into different combinations of sub-apertures, from separating each consecutive element [1 0], to each group of four elements [1 1 1 1 0 0 0 0]. In Figure 2 and Table II it can be seen that contrast is optimised, and artefacts are minimised when each consecutive element pairing [1 0] is used.

### B. In Vivo Rabbit Kidney Results

With the algorithm now optimised we then tested it in a non-contrast enhanced *in vivo* context. In Figure 4 the results can be seen imaging a rabbit kidney *in vivo* from two different angles. Qualitatively it can be seen that SAMAS offers the highest contrast and lowest artefacts, this can be confirmed by quantifying the SNR and CNR relative to locations (C, C', D, and D'). Similarly to what was shown in the *in vitro* experiments the SAMAS method also benefits more from high ensemble lengths than the other methods. In figure 4 a line cut through of a few major blood vessels can be seen, demonstrating the enhanced resolution of SAMAS and a contrast enhancement over ASAP between 8 and 13 dB. The red and black arrows highlight where SAMAS outperform the other methods thanks to the temporal and spatial processing: the red arrows highlight where noise reduction is improved thanks to the ASAP methods while the black arrows highlight where the contrast is improved thanks to FMAS method.

TABLE III
SIGNAL-TO-NOISE RATIO (SNR) AND CONTRAST-TO-NOISE RATIO (CNR) FOR FOUR ROIS AS INDICATED IN FIG 4 FOR AN ENSEMBLE LENGTH OF 600 SAMPLES. THE BOLD TYPE PROVIDES THE dB DIFFERENCE BETWEEN SAMAS AND OTHER METHODS.

|  | ROI | PD CC | PD FMAS | ASAP | SAMAS |
|---|---|---|---|---|---|
| SNR (dB) | C | 10.1 (**12.8**) | 14.1 (**8.8**) | 14.6 (**8.4**) | **22.9** |
|  | C' | 6.6 (**15.9**) | 11.8 (**10.8**) | 13.4 (**9.1**) | **22.6** |
|  | D | 18.2 (**18.9**) | 20.8 (**16.3**) | 30.8 (**6.3**) | **37.1** |
|  | D' | 11.9 (**8.9**) | 15.7 (**5.1**) | 14.7 (**6.1**) | **20.8** |
| CNR (dB) | C | 16.4 (**6.4**) | 18.6 (**4.3**) | 15.6 (**7.2**) | **22.9** |
|  | C' | 14.3 (**9.4**) | 18.4 (**5.2**) | 14.8 (**8.8**) | **23.6** |
|  | D | 25.4 (**10.6**) | 26.9 (**9.1**) | 29.7 (**6.3**) | **36.0** |
|  | D' | 15.2 (**5.7**) | 18.9 (**2.0**) | 15.3 (**5.6**) | **20.9** |

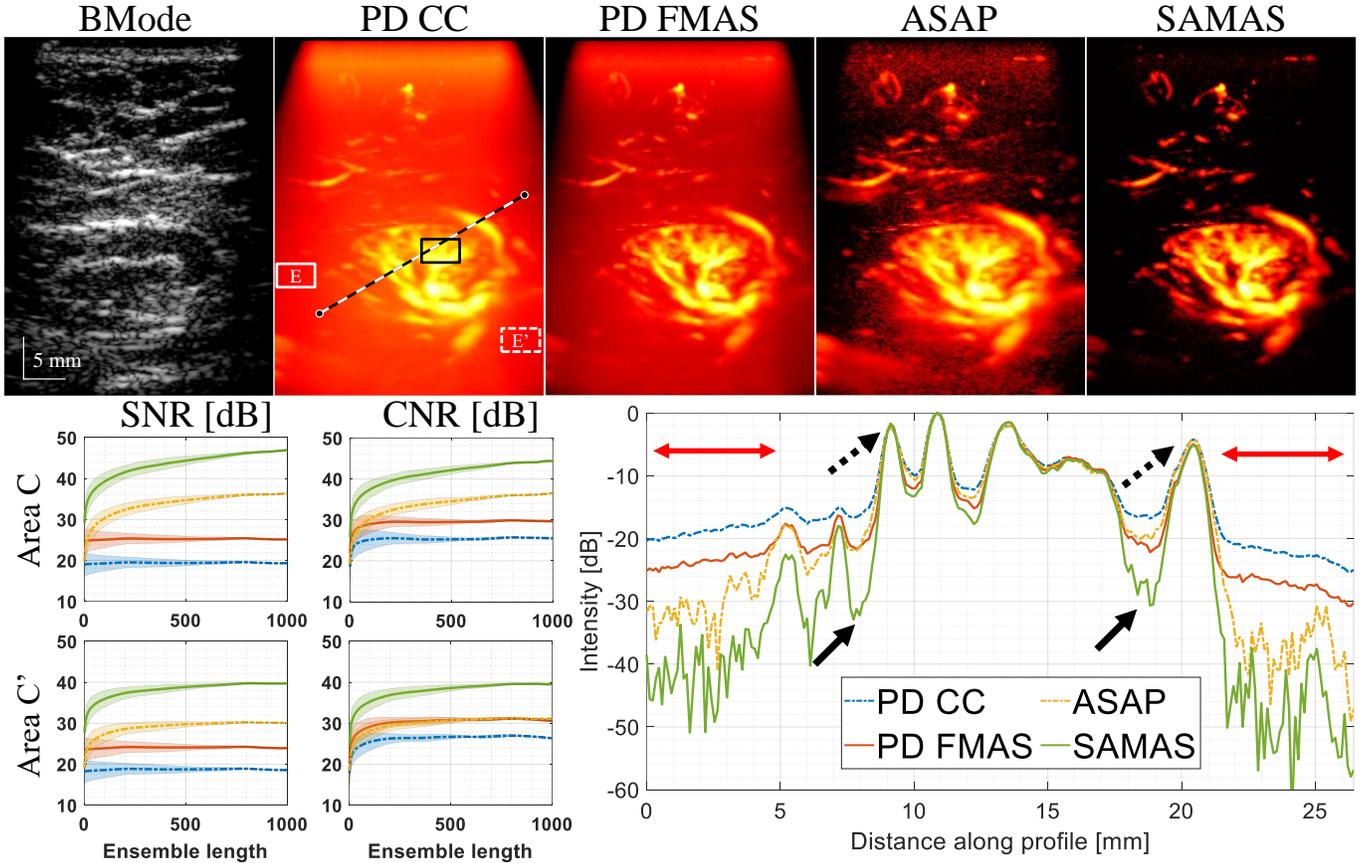

Fig. 5. BMode, PD CC, PD FMAS, ASAP, and SAMAS contrast imaging of a lymph node. The SNR and CNR over varying ensemble lengths for areas E and E' are provided as well as a cut-through line profile comparing each of the beamforming techniques. BMode and contrast images are displayed with a 45 dB dynamic range.

TABLE IV
SIGNAL-TO-NOISE RATIO (SNR), AND CONTRAST-TO-NOISE RATIO (CNR) FOR TWO ROIS AS INDICATED IN FIG 5 FOR AN ENSEMBLE LENGTH OF 1000 SAMPLES. THE BOLD TYPE PROVIDES THE dB DIFFERENCE BETWEEN SAMAS AND OTHER METHODS (POSITIVE VALUES MEANS IMPROVEMENT).

| | ROI | PD CC | PD FMAS | ASAP | SAMAS |
|---|---|---|---|---|---|
| SNR (dB) | E | 19.3 (**27.6**) | 25.1 (**21.8**) | 36.4 (**10.6**) | **47** |
| SNR (dB) | E' | 18.6 (**21.1**) | 23.9 (**15.8**) | 30.1 (**9.6**) | **39.7** |
| CNR (dB) | E | 25.5 (**18.9**) | 29.6 (**14.8**) | 36.5 (**7.9**) | **44.4** |
| CNR (dB) | E' | 26.3 (**13.2**) | 30.6 (**8.9**) | 31.1 (**8.4**) | **39.5** |

### C. In Vivo contrast Lymph Results

To explore the algorithm further we investigated a novel application. Human lymph node imaging utilising *intra-venous* contrast agents. The results of this can be seen in Figure 5 and table IV. The CNR and SNR of SAMAS is enhanced over the alternatives. Additionally, it once again can be seen that SAMAS benefits more from increasing ensemble length than the alternative methods. The line cut through shows SAMAS' ability to separate the individual blood vessels.

## IV. DISCUSSION

In this study we propose the SAMAS algorithm, which combines the advantages of the FMAS and ASAP methods. Both the incoherence of noise over time in FMAS and the incoherence of noise over space (data channels) in ASAP were utilised in SAMAS to achieve superior noise suppression. First we optimised the use of phase information and sub-aperture pairing to minimise artefacts *in vitro*. We then demonstrated the capabilities of the algorithm *in vivo*, first in a rabbit kidney and then by generating human lymph node ultrafast ultrasound images utilising *intra-venous* contrast agents.

While this study focuses on ultrasound data enhanced by using microbubble contrast agents, the proposed approach can also be applied to ultrafast ultrasound data obtained without contrast agents. Separately, FMAS and ASAP have already been applied to non-contrast applications [16], [25], and it would be interesting to explore the potential improvement of SAMAS over those.

TABLE V
COMPUTATIONAL MEAN AND STANDARD DEVIATION TIME OF 5-CAROTID BIFURCATION PHANTOM IMAGES IN MILLISECONDS PER FRAME FOR EACH TECHNIQUE.

|  | PD CC | PD FMAS | ASAP | SAMAS |
|---|---|---|---|---|
| **Loading data** | 58.6 ± 1.4 | 57.8 ± 0.3 | 78.0 ± 0.3 | 78.3 ± 0.3 |
| **Beamforming** | 84.6 ± 1.1 | 275.8 ± 2.5 | 161.5 ± 0.4 | 543.5 ± 1.4 |
| - DAS beamformer | 73.5 ± 7.2 | 80.5 ± 5.1 | 147.2 ± 16.3 | 158.2 ± 7.2 |
| - Sum or FMAS | 3.4 ± 0.4 | 188.1 ± 5.9 | 6.7 ± 0.7 | 376.9 ± 8.5 |
| - Cross / Auto correlation | 1.0 ± 0.1 | 1.0 ± 0.1 | 1.3 ± 0.1 | 2.1 ± 0.2 |
| - Ensemble averaging | 1.2 ± 13.7 | 1.1 ± 12.7 | 0.8 ± 0.1 | 0.9 ± 0.1 |
| - Sidelobe suppressor | NC | NC | 11.8 ± 0.7 µs | 13.6 ± 1.4 µs |

One limitation of this study is the lack of ground truth knowledge in the *in vivo* quantification: we cannot know for certain whether our "noise" regions are in fact pure noise or contain some signal. This is a limitation inherent to all *in vivo* experiments without a ground truth when attempting to devise quantitative parameters for analyzing different beamforming techniques.

Another limitation is the computational cost of SAMAS. Table V provides the average and standard deviation of the time taken to compute each frame when generating 5 images of the carotid bifurcation phantom with PD CC, PD FMAS, ASAP and SAMAS. The times were obtained using the *tic* and *toc* Matlab functions on a workstation with an Intel Core i9 9900K @4.7GHz, 128GB RAM @3200 MHz and a Nvidia Titan XP. Two main times are provided in the table, one for loading the data and the other for beamforming. Loading data includes the loading of the RF frames from a SSD, as well as the channel splitter for ASAP and SAMAS, and a clutter filter (Fig. 1). The subsidiary beamforming steps included in the total beamforming time are also provided. Apart from the DAS beamformer, which used a custom CUDA GPU process, all the other processes were realized in Matlab, without further optimization.

Concerning the loading time, ASAP and SAMAS were already slower than the other methods. This is because the channels have to be split, and more datasets are handled. For the beamforming, there is a large difference between CC and FMAS, the bottleneck in the FMAS process being the multiplication between all the possible angle combinations taking a long time in Matlab. That increased the time by a factor of 55. Regarding ASAP, as expected, the method takes twice as long as CC because the DAS beamformer and the sum of angles need to be performed twice. As expected, SAMAS inherits each time constraint from FMAS and ASAP. However, as demonstrated in our recent work [12], when FMAS is executed by a Cuda GPU script, it can have a similar computational time to CC. We therefore expect that the computational time for SAMAS could be reduced by a factor of 55, making it similar to that of ASAP.

The SAMAS algorithm consistently improved the CNR and SNR across all our tests. On average it improved the CNR by 11 dB and the SNR by 18 dB *in vivo*. We hope that this work lays the groundwork for a promising method which could see widespread clinical use.